\documentclass[pra,twocolumn,preprintnumbers,showpacs]{revtex4} 
\usepackage{graphicx} 
\usepackage{dcolumn} 
\usepackage{bm} 
\usepackage{amsmath}

\usepackage[dvips]{color}

\usepackage{epsfig}

\begin{document}

\title{Theory of wave-front reversal of short pulses in dynamically-tuned zero-gap periodic systems}
\author{Yonatan Sivan and John B. Pendry} \affiliation{The Blackett Laboratory, Department of Physics, Imperial College London, London SW72AZ}
\email{ysivan@imperial.ac.uk}

\begin{abstract}
Recently~\cite{Sivan-Pendry-letter}, we have shown that the wave-front of short pulses can be accurately and efficiently reversed by use of simple one-dimensional zero-gap photonic crystals. In this Article, we describe the analytical approach in detail, and discuss specific structures and modulation techniques as well as the required steps for achieving complete time-reversal. We also show that our scheme is only very weakly sensitive to material losses and dispersion.
\end{abstract}

\pacs{42.25.Bs, 42.70.Qs, 42.79.Hp}

\maketitle

\section{Introduction}\label{sec:introduction}
Time-reversal is one of the most spectacular yet elusive wave phenomena. A time-reversed pulse evolves as if time runs backward, thus eliminating any distortions or scattering that occurred at earlier times. This enables light detection, imaging and focusing through complex media (see e.g.,~\cite{Fink_review_1,Fink_review_2,Fink_review_3} and references therein) with applications in diverse fields such as medical ultrasound~\cite{Fink_review_1,Fink_review_2,Fink_review_3}, communication systems and adaptive optics~\cite{Pepper-book,magnonic_crystals_transmission_lines,Fink_microwave_reversal}, superlensing~\cite{tr_super_lens_pendry}, ultrafast plasmonics~\cite{Stockman_reversal}, biological imaging~\cite{yaqoob_Nat_Phot}, THz imaging~\cite{THz_imaging} and quantum information and computing~\cite{Cucchietti}. 

For low frequency waves (e.g., in acoustics, spin waves, elastic waves etc.), time-reversal can be accomplished by electronic sampling, recording, and playing back~\cite{Fink_review_1,Fink_review_2,Fink_review_3}. This is possible since in this frequency range, the pulse oscillates on a scale slower than electronic sampling speed. On the other hand, for high frequency waves, specifically, optical electromagnetic waves, the period of the pulse oscillations (namely, the inverse of the carrier frequency) is too fast to be sampled properly by even the fastest electronic detector. Moreover, for sufficiently short pulses, even the pulse's envelope is too short to be sampled electronically.

The standard solution in the optical regime is to use nonlinear processes such as Three-Wave Mixing or (nearly-degenerate) Four-Wave Mixing (TWM or FWM, respectively)~\cite{Pepper-book,Boyd-book}. When these techniques are applied to monochromatic (CW) beams (i.e., when the electric field of the pulse is given by $E = A(x,y,z) e^{- i \omega_0 t} + c.c.$), a process known as Phase-Conjugation (PC), the incident photon interacts with two CW pump beams of the same frequency such that the output photon has a negative frequency, namely, the conversion is $\omega_0 \to - \omega_0$. Mathematically, this is equivalent to a conjugation of the envelope $A(x,y,z)$, hence, the source of the terminology. The wave is then accurately time-reversed.

Application of PC with CW pump beams to a pulse ($E = A(x,y,z,t) e^{- i \omega_0 t} + c.c.$) yields the same frequency conversion, $\omega_0 \to - \omega_0$. In this case, however, the technique, which is more appropriately referred to as Temporal PC (TPC)~\cite{Tsang_Psaltis1}, does not lead to a perfect time-reversal. Indeed, since every frequency component is shifted by the {\em same} amount, the envelope conjugation is accompanied by an inversion of the spectral envelope with respect to the carrier frequency (i.e., $\hat{A}(\omega - \omega_0) \to \hat{A}^*(- \omega + \omega_0)$, with $\hat{A}$ representing the Fourier-transform of the pulse envelope). This process thus involves a non-zero momentum changes, which, in turn, cause phase-mismatch problems leading to lower efficiency and accuracy; eventually, the ability to reverse pulses with a broadband spectrum is limited. More simply put, the reversal achieved by TPC is imperfect since although the time envelope ($A(x,y,z,t)$) is conjugated, it is not time-reversed ($t$ not inverted to $-t$). As a consequence, while TPC compensates for effects such as group-velocity dispersion and Self-phase modulation (which are the dominant distortions experienced by pulses propagating in long-haul optical fibers), it does not correct for other effects (such as odd-order dispersion and self-steepening)~\cite{Tsang_Psaltis1}; Those become important for sufficiently short (usually sub-picosecond) pulses or near a zero group-velocity-dispersion point. Thus, in practice, TPC 
can accurately reverse only sufficiently long (and weak) pulses for which the effects which are not compensated for are anyhow negligible.

Among the few possible remedies to this problem, the most successful one is to employ Spectral Phase Conjugation (SPC), which is basically a phase-conjugation of the Fourier-transform of the pulse (i.e., $\hat{A}(\omega) \to \hat{A}^*(\omega)$. SPC was demonstrated using two cascaded TWM processes, see e.g.,~\cite{Marom_Fainman,Marom_Fainman_reply} and references therein. However, the complexity of the scheme~\cite{Marom_Fainman} makes it unlikely to evolve into a commercial product. A more recent implementation of SPC using two time-lens fiber systems and FWM in Silicon waveguides has been demonstrated for picosecond pulses propagating in optical fibers~\cite{Kuzucu_Gaeta}.

Several years ago, two additional time-reversal schemes have been suggested which rely on dynamically-tuned coupled resonators optical waveguide (CROW) structures. The first, by M. Yanik and S. Fan (see~\cite{Fan-reversal}), involves an adiabatic modulation of the index of refraction of two anti-symmetric CROWs. The second scheme, suggested by S. Longhi~\cite{Longhi-reversal}, is based on a time-gated spatial modulation of the index of refraction of the resonators. Both schemes can be implemented with linear modulators, e.g., using the electro-optical effect or carrier injection/depletion. They can be applied to pulses which are spectrally narrower than the bandwidth of the CROW. This, typically, corresponds to pulses not shorter than several picoseconds. To date, both schemes were not tested experimentally. 

Importantly, we note that both schemes~\cite{Fan-reversal,Longhi-reversal} perform {\em envelope (or wave-front) reversal} (i.e., $A(t) e^{- i \omega_0 t} \to A(-t) e^{- i \omega_0 t}$) rather than complete time-reversal, which requires also the conjugation of the temporal envelope. Following the discussion above, one can understand that a {\em hybrid} scheme, based on a combination of envelope reversal and TPC (which conjugates the temporal envelope) will yield perfect reversal, in a similar manner to~\cite{Fink_microwave_reversal}. 
However, as noted, the CROW-based envelope reversal schemes are spectrally limited, so the associated hybrid scheme can still reverse only relatively long pulses. 

In a recent study~\cite{Sivan-Pendry-letter}, we have introduced an alternative wave-front reversal scheme that requires a conceptually simple structure, namely, a dynamically-tuned zero-gap photonic crystal (PhC). When applied to optical pulses which are not shorter than several picoseconds, a $100\%$ reversal efficiency can be easily reached using linear modulators. Implementation of the scheme for shorter pulses is also possible, however, it requires a nonlinear mechanism such as Cross-Phase-Modulation and has a reduced efficiency. In our approach, there are no special phase-matching considerations that have to be taken into account, it can be implemented on-chip, and the required energies are significantly lower than with techniques which are based on a series of nonlinear wave-mixing processes. Furthermore, the scheme is, in principle, not limited to one-dimensional (i.e., optical fiber) systems, and can be implemented essentially in any wave system where the wave velocity varies periodically and can be modulated in time.

While in~\cite{Sivan-Pendry-letter} the scheme was described intuitively and demonstrated numerically, the analysis was described only briefly. This article is dedicated to the detailed description of the theoretical approach and the implementation details such as choice of materials, modulation techniques and the roles of dispersion and absorption. In Section~\ref{sec:principles} we recall the principles of broadband wave-front reversal in zero-gap periodic systems. In Sections~\ref{sec:Derivation_unidirectional} and~\ref{sec:envelopes} we derive the envelope equations and in Section~\ref{sec:weak_coupling_analysis} we show how to solve them analytically in the weak-coupling limit. Section~\ref{sec:design} provides a discussion on implementation considerations and Section~\ref{sec:numerics} describes the numerical results. 
Explicit formulae for the coefficients are derived in the Appendices.

\section{Principles of wave-front reversal in zero-gap systems}\label{sec:principles}

Wave propagation in periodic systems has been the focal point of countless studies in several branches of physics. Most prominent of these systems are crystalline solids. However, a variety of artificial analogues are also extremely well studied. Among those are as photonic crystals (PhCs)~\cite{Joannopoulos-book} in the context of electromagnetics, phononic crystals~\cite{phononic_crystals} in the context of acoustics, magnonic crystals~\cite{magnonic_crystals_3,magnonic_crystals_4} in the context of spin waves etc.. The study of such systems was almost exclusively associated with the forbidden gaps, which are energy/frequency regimes where waves cannot propagate. In particular, these gaps are the key to understanding the electronic and optical properties of solid-state systems; their artificial analogues have a variety of possible applications, ranging from low-threshold lasers, wave guiding, storing, filtering and switching~\cite{Joannopoulos-book} to medical ultrasound and nondestructive testing~\cite{phononic_crystals}.

Although the occurrence of bandgaps is generic to periodic systems, in some special cases, the gaps can have a {\em zero} width. Somewhat similarly, some materials and structures support two bands that cross symmetrically {\em without} forming a gap (see e.g. Fig.~\ref{fig:qws_dispersion}). One of the most notable of such structures is graphene~\cite{graphene} or its photonic crystal analogue~\cite{Joannopoulos-book} which is a two-dimensional periodic system with hexagonal symmetry.
Other examples are chiral metamaterials~\cite{chiral_Pendry}, transmission-line systems~\cite{crossing_transmission_lines}, bi-axial crystals~\cite{Berry} and spin systems~\cite{Rashba}. We refer to such systems with a perfectly symmetric crossing as {\em self-complementary} systems, since they support wave propagation in opposite directions with the same momentum. Earlier studies of self-complementary systems have already shown some unusual properties~\cite{graphene,chiral_Pendry,Berry,segev_conical_diffraction}. However, these systems are still relatively unexplored.

In this Article, we show how to exploit the unusual band-structure of a periodic self-complementary system, namely, a PhC with a zero-width gap, for the purpose of wave-front reversal of broadband pulses. As described above, since the main challenge is to reverse pulses of optical frequencies, in this Article, we focus on time-reversal of electromagnetic pulses.

The zero-gap system has two major advantages over a ``standard'' finite-gap system. First, while a pulse incident on a {\em finite}-gap system will mostly be reflected, a pulse incident on a {\em zero}-gap system would be almost perfectly admitted, with only frequencies overlapping distant gaps being repelled. The second advantage of the zero-gap system is that in contrast to the standard approach of time-reversal, which, as described in Section~\ref{sec:introduction}, is usually realized with a frequency conversion process based on TWM/FWM, the self-complementary system allows for a different approach, namely, via dynamic tuning of the refractive index. Indeed, dynamic tuning of PhCs has been one of the recent hot topics in PhC research, with various applications such as frequency shifting, light switching, slowing and stopping as well as many others studied theoretically and demonstrated experimentally, see, e.g.,~\cite{Reed_shock,Reed_doppler,Fan-stopping,Fan-OPN,Notomi_review,Lipson_review}. In the current context, the dynamic tuning enables the vertical transitions between the positive and negative group-velocity bands. The proximity of the crossing bands allows for {\em efficient} transfer of energy between the bands using {\em weak} and {\em slow} modulations of the wave velocity. In particular, the modulation can now be much slower than the period of the pulse oscillations so that the wave-front reversal can be realized electronically, e.g., using the electro-optic effect or carrier injection/depletion. However, the modulation should still fulfill several conditions. First, it should be much faster than the pulse duration or equivalently, the spectral content of the modulation should be much wider than that of the pulse. In such a case, the spectral content of the pulse is effectively constant, so that all the frequencies in the pulse are converted with the same efficiency. Only such {\em non-adiabatic} modulation ensures a frequency-independent frequency conversion resulting in a reversed pulse which is an accurate replica of the forward-propagating pulse. 
In that regard, we note that although the modulation can have fast spectral components, the possibility of transitions to higher bands in a zero-gap system is small, thus, enabling non-adiabatic modulation without a deterioration in performance. A second requirement is that the modulation be periodic in order to avoid any wavevector mixing, resulting in a vertical frequency-conversion, see Fig.~\ref{fig:qws_dispersion}(b). 


Following~\cite{Sivan-Pendry-letter}, here we analyze a simple one-dimensional zero-gap systems and confirm the intuitive arguments above. However, we emphasize that the ideas and techniques we use here can be employed also in other zero-gap systems, in one, two or three dimensions, as well as in other self-complementary systems~\cite{graphene,chiral_Pendry,crossing_transmission_lines,Berry,Rashba}. 

\section{Derivation of the uni-directional wave equations}\label{sec:Derivation_unidirectional}
\begin{figure}[htbp]
\centering{\includegraphics[scale=0.32]{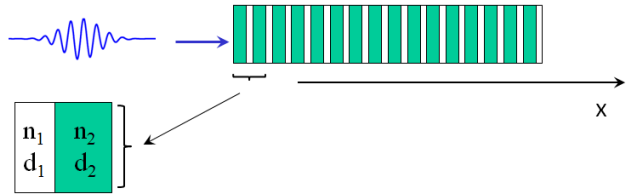}}
\caption[]{(Color online) Geometry of the pulse propagation through a layered PhC. } \label{fig:geometry}
\end{figure}

Consider an electromagnetic plane-wave pulse normally incident on a 1D Photonic Crystal (PhC) (see Fig.~\ref{fig:geometry}) which is time-modulated in the following manner
\begin{equation}\label{eq:n}
n(x,t) = n_{PhC}(x) + M_0 p(x)m(t),
\end{equation}
with $max(m(t)) = 1$ and $max(p(x)) = 1$. Here, both the static part of the refractive index, $n_{PhC}(x)$, and the spatial profile of the modulation $p(x)$ have period $d$. 
Such a periodic modulation does not introduce any wave-vector mixing, hence, allowing for an accurate reversal (see e.g.,~\cite{Fan-reversal}). The Maxwell equations in this case can be written as
\begin{eqnarray}
\nabla \times \vec{E}(x,t) &=& - \mu_0 \partial_t \vec{H}, \nonumber \\ 
\nabla \times \vec{H}(x,t) &=& \epsilon_0 \partial_t \left[n^2(x,t) \vec{E}(x,t)\right]. \nonumber
\end{eqnarray}
For a linear polarization at normal incidence,
\begin{equation}\label{eq:EH}
\vec{E} = E(x,t)\hat{z}, \quad \quad \vec{H} = H(x,t)\hat{y}, \nonumber
\end{equation}
then, the Maxwell equations reduce to
\begin{subequations}
  \label{eq:Maxwell_LP}
  \begin{eqnarray}
\partial_x E(x,t) &=& - \mu_0 \partial_t H(x,t), \\ 
\partial_x H(x,t) &=& - \epsilon_0 \partial_t \left[n^2(x,t) E(x,t)\right],
\end{eqnarray}
\end{subequations}
and eventually to
\begin{eqnarray}\label{eq:waeq}
\partial_{xx} E(x,t) - \frac{1}{c^2} \partial_{tt} \left[n^2(x,t) E(x,t)\right],
\end{eqnarray}
which is a one-dimensional wave equation with a time-dependent velocity.

We proceed with an analysis similar to that first performed in~\cite{deep_gratings_PRE_96,deep_grating_2D} in the context of pulse propagation in PhCs with a Kerr response~\footnote{We use the deep grating formulation~\cite{deep_gratings_PRE_96} because, the simpler weak grating formulation can never describe a zero-gap, even for a vanishing index contrast. }. We adopt the uni-directional field formulation by defining
\begin{subequations}
  \label{eq:EH_uni-dir}
  \begin{eqnarray}
E(x,t) &=& n^{-\frac{1}{2}}(x,t) \left[F(x,t) + B(x,t)\right], \\
Z_0 H(x,t) &=& n^{\frac{1}{2}}(x,t) \left[F(x,t) - B(x,t)\right], 
\end{eqnarray}
\end{subequations}
where $F$ and $B$ are scalar functions representing the forward and backward fluxes in the PhC~\cite{non_uniform_grating_rigorous}~\footnote{For an elaborate discussion of the accuracy of this separation for general dispersive nonlinear materials, see~\cite{Uni-directionality}. }. Substituting in Eqs.~(\ref{eq:Maxwell_LP}) gives
\begin{subequations}
  \label{eq:FB}
  \begin{eqnarray}
F_x(x,t) + \frac{n(x,t)}{c} F_t + \frac{n_t}{c} F &=& - \frac{1}{2}\left(\frac{n_t}{c} - \frac{n_x}{n}\right) B, \label{eq:F} \\
B_x(x,t) - \frac{n(x,t)}{c} B_t - \frac{n_t}{c} B &=& \frac{1}{2}\left(\frac{n_t}{c} + \frac{n_x}{n}\right) F. \label{eq:B}
\end{eqnarray}
\end{subequations}
Note that the definition of the fluxes $F$ and $B$ is valid regardless of the specific form of the refractive index distribution $n(x,t)$ and without assuming any specific modal (spatial) basis. Also note that since the wave is assumed to be normally incident on the PhC, the fields are essentially scalar and the Maxwell divergence equations are automatically satisfied. For oblique incidence, or for a higher-dimensional PhCs, the problem requires a generalized treatment, see~\cite{deep_grating_2D}.

\section{Derivation of the envelope equations} \label{sec:envelopes}
We now define
\begin{equation}\label{eq:W}
\underline{W} = \begin{pmatrix}
  F \\
  B
\end{pmatrix},
\end{equation}
so that the system~\eqref{eq:FB} can be written in matrix form as
\begin{equation} \label{eq:system}
\underline{\underline{n}}\
\underline{W}_t =
\underline{\underline{M}}\ \underline{W},
\end{equation}
where
\begin{subequations} \label{eq:nM}
\begin{eqnarray}
\underline{\underline{n}} &=& \underline{\underline{n}}_{PhC} + M_0 m(t) \underline{\underline{n}}_m  \\ 
&\equiv& \begin{pmatrix}
  n_{PhC}(x) & 0 \\
  0 & n_{PhC}(x)
\end{pmatrix} + M_0 m(t) \begin{pmatrix}
  p(x) & 0 \\
  0 & p(x)
\end{pmatrix}, \nonumber \\
 \underline{\underline{M}} &=& \begin{pmatrix}
  - c \partial_x - n_t & - \frac{c}{2}\left(\frac{n_t}{c} - \frac{n_x}{n}\right) \\
  - \frac{c}{2}\left(\frac{n_t}{c} + \frac{n_x}{n}\right) & c \partial_x - n_t
\end{pmatrix}. \label{eq:M}
\end{eqnarray}
\end{subequations}
The static index variations, $n_{PhC}$, and the dynamical variations, $p$, vary spatially on a scale comparable to $\lambda$; correspondingly, the fluxes vary in space on a similar scale. In addition, they vary in time on the scale of the period $T$. On the other hand, the pulse's envelope and the modulation $m(t)$ vary on slower scales, $T_p$ and $T_{mod}$, respectively. Thus, the solution~(\ref{eq:W}) varies on more than one temporal and spatial scale. Accordingly, we adopt a multiple-scales approach~\cite{Bender-Orszag} by assuming that the variations on the fastest scale, defined as $x_0, t_0$, correspond to variables independent from the variables describing the variations on the slower scales. However, in contrast to the standard multiple scales approach, where there are {\em only two} time scales (e.g., the period $T$ and the pulse duration $T_p$; see~\cite{Bender-Orszag} or more specifically,~\cite{deep_gratings_PRE_96}), the external modulation introduces a {\em third} independent time scale, $T_{mod}$. In order to avoid this complication to the analysis, let us lump all the slow changes into one spatial and one temporal scale, defined as $x_1$ and $t_1$, respectively. Then, Eq.~(\ref{eq:n}) becomes
\begin{equation}\label{eq:n_scales}
\underline{\underline{n}} = \underline{\underline{n}}_{PhC}(x_0) + M_0 \underline{\underline{n}}_m(x_0) m(t_1),
\end{equation}
and we can make the substitution
\begin{equation}\label{eq:multiple_scales}
\partial_t \to \partial_{t_0} + \partial_{t_1}, \quad \quad \partial_x \to \partial_{x_0} + \partial_{x_1},
\end{equation}
so that the matrix $M$ becomes
\begin{eqnarray} \label{eq:M_expansion}
\underline{\underline{M}} &=& \begin{pmatrix}
  - c \left(\partial_{x_0} + \partial_{x_1}\right) - \partial_{t_1} n  & - \frac{1}{2} \partial_{t_1} n + \frac{c}{2n} \partial_{x_0} n \\
  - \frac{1}{2} \partial_{t_1} n - \frac{c}{2n} \partial_{x_0} n  & c \left(\partial_{x_0} + \partial_{x_1}\right) - \partial_{t_1} n
\end{pmatrix}. \nonumber \\
\end{eqnarray}
In order to complete the scale separation, let us rewrite the terms involving the spatial derivatives in $\underline{\underline{M}}$~(\ref{eq:M_expansion}) as a sum of a static and dynamic parts. Using Eq.~(\ref{eq:n_scales}), it can be shown that
\begin{eqnarray}
\frac{1}{n} \partial_{x_0} n &=& \frac{n_{PhC}'(x_0)}{n_{PhC}(x_0)} \\ 
&+& \frac{ M_0 m(t_1) \left(n_{PhC} p'(x_0) - n'_{PhC} p(x_0) \right)}{n_{PhC}(x_0)\left[n_{PhC}(x_0) + M_0 p(x_0) m(t_1)\right]}, \nonumber
\end{eqnarray}
where the sign $'$ stands for a derivative with respect to the argument, a notation used only when it is unambiguous.

We can now separate $\underline{\underline{M}}$ into the different order contributions as
\begin{eqnarray}
\underline{\underline{M}} &\equiv& \underline{\underline{M}}^{(0)} + \underline{\underline{M}}^{(1)},
\end{eqnarray}
where 
\begin{subequations}
\begin{eqnarray}
\underline{\underline{M}}^{(0)}(x_0) &=& c \begin{pmatrix}
  - \partial_{x_0} & \frac{1}{2} \frac{n_{PhC}'}{n_{PhC}} \\
  - \frac{1}{2}\frac{n_{PhC}'}{n_{PhC}} & \partial_{x_0}
\end{pmatrix}, 
\\
\underline{\underline{M}}^{(1)}(x_1,t_1;x_0) &=& M_0 \frac{c}{2} \begin{pmatrix}
  0 & \beta(x_0, t_1) \\
  - \beta(x_0, t_1) & 0
\end{pmatrix} \\
&-& \underline{\underline{V}} \partial_{x_1} - M_0 \underline{\underline{C}}^T m'(t_1) p(x_0), \nonumber 
\end{eqnarray}
\end{subequations}
and where we have defined the space and time-dependent function
\begin{equation}\label{eq:beta}
\beta(x_0, t_1) = m(t_1) \frac{n_{PhC}(x_0) p'(x_0) - n'_{PhC} p(x_0) }{n_{PhC}(x_0)\left[n_{PhC}(x_0) + M_0 p(x_0) m(t_1)\right]},
\end{equation}
and the constant matrices
\begin{equation}\label{eq:V}
\underline{\underline{V}} = \begin{pmatrix} c & 0 \\
  0 & - c
\end{pmatrix}, \nonumber
\end{equation}
and
\begin{equation}\label{eq:C_T}
\underline{\underline{C}}_T = \begin{pmatrix}
  1 & \frac{1}{2} \\
  \frac{1}{2} & 1
\end{pmatrix}. \nonumber
\end{equation}
The matrix $\underline{\underline{M}}^{(0)}$ is the operator of the unperturbed system (i.e., in the absence of time-modulation) while the matrix $\underline{\underline{M}}^{(1)}$, together with the modulation term on the left-hand-side of Eq.~(\ref{eq:system}) (or equivalently, the second term in Eq.~(\ref{eq:n})), describe the slower dynamics, i.e., the evolution of the pulse envelope (wave-front) and the effects of the modulation; the dependence of $\underline{\underline{M}}^{(1)}$ on the fast scale $x_0$ is removed by integration, see below.

The formulation presented so far is valid for any 1D periodic optical systems (or PhCs). In what follows, we limit the discussion to periodic systems in which two bands have minimal interaction, leading to a gap with effective {\em zero} width, or effectively, a crossing of the two bands. Such crossing points typically appear between the higher bands of PhC structures. In most cases, they are not perfectly symmetrical with respect to the crossing point, however, nearly- or even perfectly-symmetric crossings {\em can} be found. Probably the simplest example of a {\em perfectly symmetric zero-gap} periodic system is a 1D {\em layered} PhC for which the indices and thicknesses of the two layers satisfy 
\begin{equation}\label{eq:qws_cond}
n_1 d_1 = n_2 d_2 / s, \quad \quad s = 1,2,3,\dots .
\end{equation}
In such systems, the $s + 1$ gap at vacuum wavelength~\footnote{For $s$ odd (even), the gap occurs at zero-FB momentum, $K d = K_c d = 0$ (at the edge of the Brillouin zone, $K d = K_c d = \pi$). Zero-gaps occur also at higher frequencies. }
\begin{equation}\label{eq:lambda_cr}
\lambda_c = 2 n_1 d_1 = 2 n_2 d_2 / s, 
\end{equation}
has exactly {\em zero} width. For dispersive materials, i.e., when the refractive indices depend on the wavelength, the QWS condition~(\ref{eq:qws_cond}) has to be satisfied exactly at the crossing point in order for the gap to maintain a zero-width. For simplicity, we assume that the dispersion is negligible, and defer the discussion of the role of dispersion to Section~\ref{subsec:dispersion}.

\begin{figure}[htbp]
\centering{\includegraphics[scale=0.76]{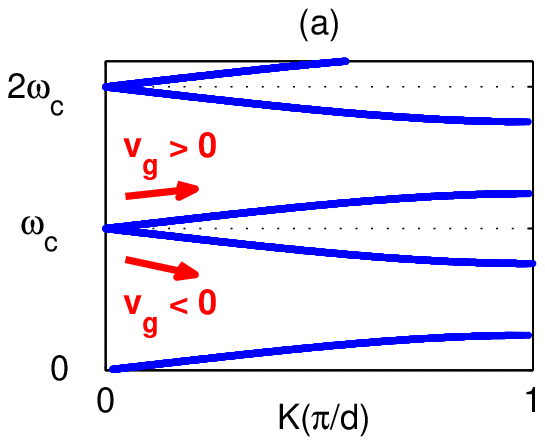} \includegraphics[scale=0.76]{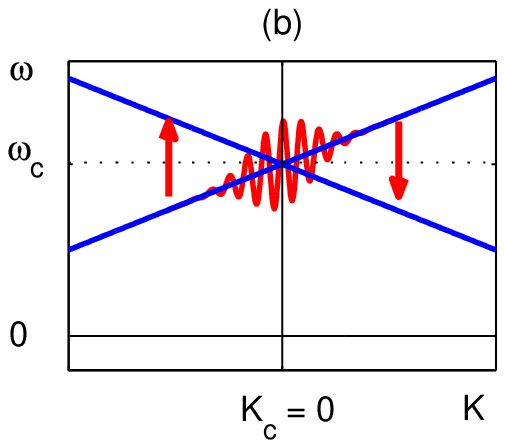}}
\caption[]{(Color online) (a) Dispersion relations of a QWS. (b) Schematic diagram of the reversal in the frequency domain. Only the part of the dispersion curve close to the crossing point $K_c = 0$ is shown. } \label{fig:qws_dispersion}
\end{figure}


When $s = 1$, this system is called a quarter-wave stack (QWS) PhC (see e.g.,~\cite{Joannopoulos-book,Yeh-book}). For simplicity, in what follows, we analyze the case of a zero-gap in the QWS system, see Fig.~\ref{fig:qws_dispersion}(b). 
As shown in Appendix~\ref{app:CMT_coeffs}, PhCs with other values of $s$ yield the same coefficients, hence, the same reversal efficiencies.

Near a gap, the solution of Eq.~(\ref{eq:system}) consists predominantly of two spectral components, one on each of the crossing bands. Each can be written as the product of a carrier uni-directional Floquet-Bloch (FB) mode $\Psi_{f/b}$ on either the positive ($f$) or negative ($b$) group-velocity band with a slowly varying envelope $f/b$, i.e.,
\begin{eqnarray}\label{eq:ansatz}
\underline{W} &=& \left[f(x_1,t_1) \underline{\Psi}_f(x_0) + b(x_1,t_1) \underline{\Psi}_b(x_0)\right] e^{- i \omega_c t_0} + c.c., \nonumber \\
\end{eqnarray}
where $\omega_c = 2 \pi c / \lambda_c$ is the band crossing frequency and $c.c.$ stands for the complex conjugate. The uni-directional FB modes $\underline{\Psi}_a$ are the eigensolutions of Eq.~(\ref{eq:system}), i.e., they satisfy
\begin{equation}\label{eq:leading_order}
- i \omega_a n_{PhC}(x_0) \underline{\Psi}_a(x_0) = \underline{\underline{M}}^{(0)}(x_0) \underline{\Psi}_a,
\end{equation}
where $\omega_a = \omega_{f/b}$ is the central frequency of the pulses on each band, chosen at the same FB wavevector.

In what follows, we neglect the complex conjugate terms in Eq.~(\ref{eq:ansatz}). This is justified for pulses which are sufficiently narrow spectrally, i.e., for pulses for which the separation of scales~(\ref{eq:multiple_scales}) is meaningful. We also ignore the population of any other distant band~\cite{deep_gratings_PRE_96}. Indeed, as shown in~\cite{deep_gratings_PRE_96}, the population of distant bands (or companion components) is manifested by the appearance of a non-zero second order term in the $k \cdot p$ expansion of the band structure. This term accounts for the group-velocity dispersion (GVD) whose magnitude scales inversely with the temporal duration of the pulse. Since, however, the QWS system exhibits strictly linear dispersion in the vicinity of the band crossing, GVD vanishes, so that population of distant bands does not play a role in the dynamics even for very short pulses~\footnote{See also discussion after Eq.~(\ref{eq:vg}). }.

The central frequency of the forward propagating input pulse can be detuned from the crossing point ($K_c = 0, \omega_c$). Since the band structure is symmetric with respect to the crossing point (see Appendix~\ref{app:FB_modes} and Fig.~\ref{fig:qws_dispersion}), and since the frequency conversion is vertical (see Fig.~\ref{fig:qws_dispersion}(b) and discussion after Eq.~(\ref{eq:vg})), we can define
\begin{equation}\label{eq:d_omega}
\delta \omega_f \equiv \omega_c - \omega_f = - \delta \omega_b = \frac{\Delta}{2}.
\end{equation}
We also assume that the detuning is small compared with the wave period, $\Delta \ll 2\pi/T$, in agreement with the above assumptions on the separation of scales~(\ref{eq:multiple_scales}).

Substituting Eq.~(\ref{eq:ansatz}) in Eq.~(\ref{eq:system}), multiplying by $\Psi_f^*$ and $\Psi_b^*$, respectively, and integrating over $x_0$ allows us to remove the dependence on the fast scale $x_0$. In particular, this leads to the following equations for the time evolution of the forward/backward envelope components
\begin{widetext}
\begin{subequations}
\begin{eqnarray} \label{eq:f}
\left(1 + M_0 m(t_1) m^{(0)}_{ff}\right) f_{t_1}(x_1,t_1) &+& v_{ff} f_{x_1} - \left(i \omega_c M_0 m^{(0)}_{ff} m(t_1) - M_0 m' m^{(1)}_{ff} + i \delta \omega_f + M_0 \beta_{ff}(t_1)\right) f \\
&=& - M_0 m^{(0)}_{fb} m(t_1) b_{t_1} - v_{fb} b_{x_1} + M_0 \left(i \omega_c m^{(0)}_{fb} m(t_1) - m' m^{(1)}_{fb} + \beta_{fb}(t_1)\right) b, \nonumber \\
\left(1 + M_0 m(t_1) m^{(0)}_{bb} \right) b_{t_1}(x_1,t_1) &+& v_{bb} b_{x_1} - \left(i \omega_c M_0 m^{(0)}_{bb} m(t_1) - M_0 m' m^{(1)}_{bb} + i \delta \omega_b + M_0 \beta_{bb}(t_1)\right) b \label{eq:b} \\
&=& - M_0 m^{(0)}_{bf} m(t_1) f_{t_1} - v_{bf} f_{x_1} + M_0 \left(i \omega_c m^{(0)}_{bf} m(t_1) - m' m^{(1)}_{bf} + \beta_{bf}(t_1)\right) f, \nonumber 
\end{eqnarray}
\end{subequations}
\end{widetext}
where we defined the coefficients as the integrals over the uni-directional FB modes, as follows,
\begin{subequations}
\begin{eqnarray}
v_{ab} &=& \int_0^d \underline{\Psi}^\dagger_a(x_0) \underline{\underline{V}}\ \underline{\Psi}_b(x_0) dx_0, \label{eq:v_ab} \\
m^{(0)}_{ab} &=& \int_0^d \underline{\Psi}^\dagger_a(x_0) \underline{\underline{n}}_m(x_0) \underline{\Psi}_b(x_0) dx_0, \label{eq:m0_ab} \\
m^{(1)}_{ab} &=& \int_0^d \underline{\Psi}^\dagger_a(x_0) \underline{\underline{n}}_m(x_0) \underline{\underline{C}}^T \underline{\Psi}_b(x_0) dx_0, \label{eq:m1_ab} \\
\beta_{ab}(t_1) &=& \frac{c}{2} \int_0^d \underline{\Psi}^\dagger_a(x_0) \underline{\underline{\beta}}(x_0,t_1)\  \underline{\Psi}_b(x_0) dx_0. \label{eq:beta0_ab}
\end{eqnarray}
\end{subequations}
Eqs.~(\ref{eq:f})-(\ref{eq:b}) are transport equations with time-dependent velocities, coupled by the modulation $m(t_1)$. The equations for $f$ and $b$ are symmetrical, however, the initial condition is highly asymmetric, consisting exclusively of forward waves (see also Section~\ref{sec:numerics}). Moreover, note that the form of the coefficient of the dynamic terms $f_{t_1}$ and $b_{t_1}$ (namely, $1 + M_0 ...$) implies that a modulation that increases the refractive index ($M_0 > 0$) will not give the same results as a modulation that decreases the refractive index ($M_0 < 0$). This effect, however, is significant only once the change in the pulse velocity due to the modulation is appreciable.

The existence of simple explicit analytical formulas for the FB modes of the layered system (see e.g.,~\cite{Yeh-book}), allows us to compute the coefficients~(\ref{eq:v_ab})-(\ref{eq:m1_ab}) analytically. Near the band crossing, the obtained expressions are relatively simple, and shed light on the physical meaning of the various terms in Eqs.~(\ref{eq:f})-(\ref{eq:b}). First, the diagonal velocity coefficients $v_{ff}$ ($v_{bb}$) represent the velocity of the forward (backward) pulses. Indeed, it is shown in~\cite{deep_gratings_PRE_96} that the $v_{aa}$ term is equal to the first term in the $k \cdot p$ expansion of the band structure, i.e., to the group-velocity; also, in Appendix~\ref{app:v} it is shown that
\begin{equation}\label{eq:vg}
v_{ff} = - v_{bb} \equiv v_g \equiv \frac{\partial \omega}{\partial k}\Big|_{\omega_c} = \frac{c}{\sqrt{n_1 n_2}}.
\end{equation}
The off-diagonal velocity coefficients $v_{ab}$ are responsible for reflection coupling (i.e., for horizontal transitions $k \to -k$)~\cite{deep_gratings_PRE_96}. Intuitively, the strictly linear dispersion relations at the crossing point imply that no reflection coupling is taking place, i.e., that $v_{bf} = v_{fb} = 0$. Indeed, this is confirmed in Appendix~\ref{app:v}~\footnote{Indeed, the rigorous relation between the band structure curvature (GVD) and $v_{bf}$ is given in~\cite{deep_gratings_PRE_96}. }. The vanishing of the {\em reflection} coupling ensures that the modulation-induced transitions between the bands are strictly vertical so that the frequency conversion is free of wavevector mixing, see Fig.~\ref{fig:qws_dispersion}(b). 

The diagonal $m_{aa}^{(0)}$ coefficients describe the phase accumulation induced by the modulation. Indeed, it is shown in Appendix~\ref{app:m0} that these diagonal coefficients are identical, i.e., $m_d \equiv m^{(0)}_{ff} = m^{(0)}_{bb}$, and that for a piecewise-uniform modulation, they are given by
\begin{equation}\label{eq:md}
m_d = \frac{1}{2}\left(\frac{n_{m1}}{n_1} + \frac{n_{m2}}{n_2}\right), 
\end{equation}
where $n_{mj}$ is the amount by which layer $j$ is modulated~\footnote{Typically, $n_{mj} = 1$ ($n_{mj} = 0$) for a layer which is modulated (not modulated)). These coefficients become spatially-dependent for a non-uniform modulation. }. 

The diagonal $m_{aa}^{(1)}$ coefficients represent the change of the electromagnetic energy induced by the modulation. In Appendix~\ref{app:m1} it is shown that for a piecewise-uniform modulation, the coefficients $m_{aa}^{(0)}$ and $m_{aa}^{(1)}$ are equal. 

The off-diagonal coefficients $m_{fb}^{(j)} = {m_{bf}^{(j)}}^*$ are responsible for {\em reversal} coupling, i.e., to the vertical transitions $\omega + \delta \omega \to \omega - \delta \omega$, see Fig.~\ref{fig:qws_dispersion}(b). One can interpret these coefficients as the measure of the imperfect spatial overlap (or equivalently, the phase mismatch) between the modes immediately above and below the crossing frequency. Their mathematical form (see Eqs.~(\ref{eq:m0_ab})-(\ref{eq:m1_ab})), which is reminiscent of the orthogonality integral (see e.g., Eq.~(\ref{eq:orthogonality_phi})), implies that their size should increase with the index contrast between the layers. Indeed, it is shown in Appendices~\ref{app:m0}-\ref{app:m1} that for a piecewise-uniform modulation and zero-detuning, the coupling coefficients are given by
\begin{equation}\label{eq:m_od}
m_{od} = m^{(0)}_{fb} = m^{(1)}_{fb} = \frac{1}{2} \frac{n_2 - n_1}{n_2 + n_1}\left(\frac{n_{1m}}{n_1} - \frac{n_{2m}}{n_2}\right).
\end{equation}
Thus, the reversal coupling tends to zero as the index contrast approaches zero or if the modulation maintains the QWS condition~(\ref{eq:qws_cond}), i.e., if no gap is opened due to the modulation. In all other cases, the dependence of the coupling coefficient on the indices is non-trivial, see also Section~\ref{subsec:materials}. It can also be shown that for non-zero detuning, $m_{od}$ becomes complex.


Lastly, the $\beta_{ab}$ coefficients~(\ref{eq:beta0_ab}) are particularly intriguing. They are defined in terms of an integration over $\beta$~(\ref{eq:beta}). For the layered structure under consideration, $n'_{PhC}$ (and $n'_m$ as well for a piecewise continuous modulation) vanishes everywhere except at the interfaces between the layers, where it becomes infinite. In these cases, the coefficients~(\ref{eq:beta0_ab}) can be evaluated by smoothing the index discontinuities over a finite length $L$, computing the FB modes numerically, and then taking the limit $L \to 0$. Our numerical computations, performed for several configurations of indices, thicknesses, and sizes of time-localized modulations, indicate that for a layered PhC, these terms vanish, $\beta_{ab}(L \to 0) = 0$. Nevertheless, any other configuration (namely, a smooth index PhC, spatially non-uniform modulations etc.) requires the specific numerical evaluation of $\beta_{ab}$. 

Due to all the above, for a piecewise-uniform modulation, Eqs.~(\ref{eq:f})-(\ref{eq:b}) simplify into
\begin{widetext}
\begin{subequations}
\begin{eqnarray} \label{eq:f_uniform}
\left(1 + M_0 m_d m(t_1)\right) f_{t_1}(x_1,t_1) &+& v_g f_{x_1} + M_0 m_d\left(m' - i \omega_c m(t_1)\right) f - i \delta \omega_f f \nonumber \\
&=& M_0 m^{(0)}_{od} \left(i \omega_c m(t_1) b - m' b - m(t_1) b_{t_1}\right), \\
\left(1 + M_0 m_d m(t_1)\right) b_{t_1}(x_1,t_1) &-& v_g b_{x_1} + M_0 m_d\left(m' - i \omega_c m(t_1) \right) b - i \delta \omega_b b \nonumber \\
&=& M_0 m_{od}^* \left(i \omega_c m(t_1) f - m' f - m(t_1) f_{t_1}\right). \label{eq:b_uniform}
\end{eqnarray}
\end{subequations}
\end{widetext}
In what follows, we focus on this set of equations.

Before finishing this Section, we would like to comment on alternative derivations of envelope equations for pulse (or beam) propagation in periodic structures. First, similar equations to Eqs.~(\ref{eq:f_uniform})-(\ref{eq:b_uniform}) can be derived using the Slowly-Varying Envelope Approximation (SVEA), i.e., by assuming that the carrier wave is a Fourier mode (i.e., a plane-wave) rather than a Floquet-Bloch (FB) mode. However, not only the uni-directional wave formulation is simpler, more elegant and more accurate than a derivation base on the SVEA~\cite{Sipe_Salinas_CMT_deep_gratings_OC}, but for PhCs, the SVEA is not strictly valid, since in the presence of $\lambda$-fast index variations, the second derivatives are not small compared to the other terms. An exception is, possibly, the case in which the index contrast is small (or in other words, that the grating is shallow), see e.g.~\cite{non_uniform_grating_rigorous}. Indeed, in that case, the difference between the Fourier and FB bases is small.

A different approach described by Craster {\em et al.} is to derive envelope equations for the wave equation (rather than for the directional flux equations, as in our derivation). This approach was employed to a large variety of periodic structures in one and two dimensions, see e.g.,~\cite{Craster_P_ROY_SOC}.

Finally, a different set of carrier modes, computed from the transmission and reflection data of a given {\em finite size} PhCs, were used in coupled Helmholtz equations in the context of three quasi-monochromatic wave mixing in PhCs, see~\cite{review_nlo_in_PhC} and references therein. However, this approach is only suitable for pulses much longer than the PhC, since in that case, the reflections from the ends of the PhC system are important.

\section{Weak coupling analysis}\label{sec:weak_coupling_analysis}
In principle, Eqs.~(\ref{eq:f_uniform})-(\ref{eq:b_uniform}) are significantly simpler to solve numerically compared with the wave equation~(\ref{eq:waeq}) since they do not require to resolve the $\lambda-$ and $T-$scale oscillations. In practice, however, the numerical solution of the envelope equations is non-trivial (see Section~\ref{sec:numerics} for details); Moreover, they are also difficult to solve analytically. In order to facilitate the analysis, recall that in our case, only the forward component (i.e., the positive group-velocity band) is initially populated. Thus, as long as the backward component is small compared with the forward component, we can neglect all the coupling terms in the equation for the forward wave~(\ref{eq:f_uniform}), i.e., we can set its right-hand-side to zero~\footnote{This assumption is reminiscent of the ``un-depleted pump'' approximation, which is common for wave mixing processes, see e.g.,~\cite{Boyd-book}. However, in the current context, it is the signal (i.e., forward) wave that is assumed to be ``un-depleted'' rather than the pump (i.e. the modulation, which is external in our case). }. In other words, we assume that there is only  forward-to-backward coupling, or simply, that the coupling is weak. In this case, a closed form solution can be obtained to Eqs.~(\ref{eq:f_uniform})-(\ref{eq:b_uniform}). 

In order to derive that solution, note that the characteristics of Eqs.~(\ref{eq:f_uniform})-(\ref{eq:b_uniform}) are solutions of 
\begin{equation}\label{eq:characteristics_full_f}
dx_1 = \pm \frac{v_g dt_1}{1 + M_0 m_d m(t_1)}. \nonumber 
\end{equation}
We integrate and define
\begin{equation}\label{eq:moving_frames_transform}
x_1^{(f,b)} \equiv x_1 \mp v_g \int_{-\infty}^{t_1} \frac{dt_1'}{1 + M_0 m_d m(t'_1)}. \nonumber
\end{equation}
This is equivalent to transforming into frames moving with each of the pulses. Then, Eqs.~(\ref{eq:f_uniform})-(\ref{eq:b_uniform}) reduce to
\begin{widetext}
\begin{eqnarray}
\left(1 + M_0 m_d m(t_1)\right) f_{t_1}\left(x_1^{(f)},t_1\right) &=& \left(i \omega_c M_0 m(t_1) m_d + i \delta \omega_f - M_0 m_d m'\right) f, \label{eq:f_final_full} \\
\left(1 + M_0 m_d m(t_1)\right) b_{t_1}\left(x_1^{(b)},t_1\right) &=& \left(i \omega_c M_0 m(t_1) m_d + i \delta \omega_b - M_0 m_d m'\right) b \label{eq:b_final_full} \\
&-& M_0 m(t_1) m_{od} \left(f_{t_1} - i \omega_c f\right) - M_0 m' m_{od} f\left(x_1^{(b)} - 2 v_g \int_{-\infty}^{t_1} \frac{dt_1'}{1 + M_0 m_d m(t'_1)}, t_1\right). \nonumber
\end{eqnarray}
\end{widetext}
We now define $\tau = \left(M_0 m_d \omega_c\right)^{-1}$. The solution of Eq.~(\ref{eq:f_final_full}) can be written as
\begin{eqnarray}
f\left(x_1^{(f)},t_1\right) &=& e^{\int_{-\infty}^{t_1} p_f(t_1') dt_1'} f_0\left(x_1^{(f)}\right), \nonumber \\ 
p_f(t_1') &=& \frac{i \frac{m(t_1')}{\tau} + i \delta \omega_f - M_0 m_d m'(t'_1)}{1 + M_0 m_d m(t_1')}, \label{eq:f_sol_full}
\end{eqnarray}
where $f_0$ is the envelope of the pulse traveling in the forward direction before the onset of the modulation. Similarly, we define
\begin{eqnarray}\label{eq:b_bar}
b\left(x_1^{(b)},t_1\right) &=& e^{\int_{-\infty}^{t_1} p_b(t_1') dt_1'} \bar{b}\left(x_1^{(b)},t_1\right), \nonumber \\ 
p_b(t_1') &=& \frac{i \frac{m(t'_1)}{\tau} + i \delta \omega_b - M_0 m_d m'(t'_1)}{1 + M_0 m_d m(t'_1)}.
\end{eqnarray}
Note that for zero detuning and uniform modulation, $p_b = p_f$. Also note that the definitions of $p_f$ and $p_b$ confirm the physical interpretation of the $m_d$ coefficients~(\ref{eq:md}), namely, the phase accumulation and change of electromagnetic energy due to the modulation, as described in Section~\ref{sec:envelopes}. Hence,
\begin{widetext}
\begin{eqnarray}
\bar{b}\left(x_1^{(b)},t_1\right) &=& - M_0 m_{od} \int_{-\infty}^{t_1} \frac{m(t'_1) f_{t'_1} - i \omega_c m(t'_1) f + m'(t'_1) f}{1 + M_0 m_d m(t'_1)} e^{- \int_{-\infty}^{t_1'} p_b(t_1'') dt_1''} dt_1'. \label{eq:b_bar_final_full}
\end{eqnarray}
We substitute Eqs.~(\ref{eq:f_final_full}) and~(\ref{eq:f_sol_full}) into Eq.~(\ref{eq:b_bar_final_full}) and get after some algebra
\begin{eqnarray}
\bar{b}\left(x_1^{(b)},t_1\right) &=& - M_0 m_{od} \int_{-\infty}^{t_1} h(t_1'; M_0, \omega_f) e^{i \Delta t_1'} f_0\left(x_1^{(b)} - 2 v_g \int_{-\infty}^{t_1'} \frac{dt_1''}{1 + M_0 m_d m(t''_1)}\right) dt_1', \nonumber \\ \label{eq:b_sol_full}
\end{eqnarray}
\end{widetext}
where
\begin{equation}\label{eq:impulse_response}
h(t_1'; M_0, \omega_f) = \frac{- i \omega_f m(t'_1) + m'(t'_1)}{\left[1 + M_0 m_d m(t'_1) \right]^2}.
\end{equation}
At times long after the modulation has ended we can set the upper limit of both integrations to $\infty$. Thus, Eq.~(\ref{eq:b_sol_full}) shows that the backward wave is given by a convolution of the forward wave 
with $h(t_1'; M_0, \omega_f)$, which has the role of the impulse response of the system. This mathematical form of the backward wave is reminiscent to the results of wave-mixing-based spectral phase conjugation~\cite{Marom_Fainman} in which the backward wave is a given by a convolution of the signal with both pump waves. 

It follows from Eqs.~(\ref{eq:b_sol_full})-(\ref{eq:impulse_response}) that {\em accurate reversal can be achieved only in the non-adiabatic limit}, i.e., when
\begin{equation} \label{eq:non-adiabatic}
T_{mod} \ll T_p,
\end{equation}
in which case the impulse function is essentially a delta function.

As an example, let us assume a Gaussian modulation
\begin{equation}\label{eq:gaussian_mod}
m\left(t\right) = e^{- \frac{(t - t_0)^2}{T_{mod}^2}}.
\end{equation}
For such a symmetric modulation, one expects the contribution of the $m'(t)$ term  in $h$~(\ref{eq:impulse_response}) to be small compared with the contribution from the $\omega_f m(t)$ term. Indeed, first, $m'(t) \sim T_{mod}^{-1} \ll \omega_f$. Second, $m'(t)$ has odd symmetry, so that contributions from times before and after the middle of the modulation $t_0$ nearly cancel. This was confirmed with asymptotic calculations (not shown). Thus, in what follows, we neglect the $m'(t)$ term in $h$~(\ref{eq:impulse_response}).

For a small modulation, we can also neglect the $M_0$ term from the denominators, i.e., we effectively assume that the pulses' velocities are not affected by the modulation. In this case, $x^{(f,b)}_1 = x_1 \mp v_g t_1$. Using Eq.~(\ref{eq:b_bar}), it then can be shown that Eq.~(\ref{eq:b_sol_full}) reduces to
\begin{widetext}\begin{eqnarray}
b\left(x_1^{(b)},t_1\right) &\cong& i \omega_f M_0 m_{od} e^{i \delta \omega_b t_1 - M_0 m_d m(t'_1) + \frac{i}{\tau} \int_{-\infty}^{t_1} m(t'_1)} \int_{-\infty}^\infty m(t'_1) e^{i \Delta t_1'} f_0 \left(x_1^{(b)} - 2 v_g t_1'\right) dt_1'. \nonumber \\
\label{eq:b_sol_small}
\end{eqnarray}\end{widetext}
This integral can be solved exactly for a Gaussian pulse 
\begin{equation}\label{eq:f_inc_gaussian}
f_0\left(\frac{x_1}{v_g T_p}\right) = e^{- \frac{x_1^2}{v_g^2 T_p^2}},
\end{equation}
where $T_p \cong 0.7 W/v_g$, with $W$ being the full-width at half-maximum (FWHM) of the incident pulse. In this case, the convolution integral is
\begin{eqnarray}
I_C &\equiv& \int_{-\infty}^\infty e^{i \Delta t_1'} e^{- \frac{(t'_1 - t_0)^2}{T_{mod}^2}} e^{- \frac{{x_1^{(b)}}^2 - 4 v_g t_1' x_1^{(b)} + 4 v_g^2 {t_1'}^2}{v_g^2 T_p^2}} dt_1', \nonumber \\
\end{eqnarray}
with
\begin{equation}\label{eq:Teff}
\frac{1}{T_{eff}^2} = \frac{1}{T_{mod}^2} + \frac{4}{T_p^2}.
\end{equation}
Then, it can be shown that
\begin{eqnarray}
I_C &=& \sqrt{\pi} T_{eff} f_0\left(\frac{{x_1^{(b)}} - 2 v_g t_0}{v_g \sqrt{T_p^2 + 4 T_{mod}^2}}\right) e^{- \frac{\Delta^2}{4} T_{eff}^2} e^{i\Phi}, \nonumber
\end{eqnarray}
with $\Phi = \frac{i \Delta}{T_p^2 + 4 T_{mod}^2} \left(T_p^2 t_0 + 2 T_{mod}^2 \frac{x_1^{(b)}}{v_g}\right)$. This shows that the envelope of the backward pulse is indeed a properly reversed version of the envelope of the forward pulse; however, it is also broader than the forward pulse due to the convolution with the modulation~\footnote{There is also a spatio-temporal frequency shift, occurring for non-zero detuning. This term is negligible in the non-adiabatic limit. We neglect it below. }. More generally, arbitrary asymmetric pulses will also undergo some distortion due to that convolution. These effects can be minimized in the non-adiabatic limit~(\ref{eq:non-adiabatic}), where we get
\begin{eqnarray}
I_C &=& \sqrt{\pi} T_{mod} f_0\left(\frac{x_1^{(b)} - 2 v_g t_0}{v_g T_p}\right) e^{- \frac{\Delta^2}{4} T_{mod}^2} e^{i \Delta t_0}. \nonumber \\
\end{eqnarray}
Finally, substituting in Eq.~(\ref{eq:b_sol_small}) gives an overall reversal amplitude of
\begin{eqnarray}\label{eq:abs_b_weak_small_m0}
max_{t_1}|b(x_1,t_1)| = \sqrt{\pi} T_{mod} \omega_f M_0 |m_{od}| e^{- \frac{\Delta^2 T_{mod}^2}{4}} f_0(0). \nonumber \\
\end{eqnarray}
Note that the solution is symmetric with respect to the sign of $M_0$, because we neglected the effect of the modulation on the pulse velocity. It also shows an exponential (Gaussian) decrease of the efficiency with the detuning.

The calculation above can be generalized for arbitrarily-shaped pulses. Moreover, the calculation can be performed without the simplifications applied to the impulse response~(\ref{eq:impulse_response}). Inclusion of these terms yields $O(M_0^2)$ corrections to the amplitude of the reversed pulse. However, such a calculation does not take into account even $O(M_0)$ corrections to the amplitude and profile of the {\em forward} wave. This can be amended by self-consistency iterations, but this would be tedious and would probably extend the validity of the perturbation analysis only slightly. More generally, once the backward wave becomes comparable in magnitude to the forward wave, in particular, for sufficiently long modulation times and/or strong coupling, the weak coupling solution~(\ref{eq:abs_b_weak_small_m0}) is not valid anymore. In all those cases one needs to solve the envelope equations~(\ref{eq:f_uniform})-(\ref{eq:b_uniform}) {\em without} approximation. Nevertheless, despite these limitations, we show in Section~\ref{sec:numerics} that the weak coupling solution~(\ref{eq:abs_b_weak_small_m0}) provides a very good approximation to the actual dynamics (obtained by solving the wave Eq.~(\ref{eq:waeq}) and the envelope equations~(\ref{eq:f_uniform})-(\ref{eq:b_uniform})) even for conversion efficiencies as high as $50\%$.

\section{Design rules}\label{sec:design}

We now turn to discuss the practical aspects of designing our Reversal Mirror (RM) for electromagnetic pulses. We take into account realistic material parameters, discuss efficiencies and a variety of other practical issues.


\subsection{RM thickness}
The length of the RM, $D$, should be chosen such that it contains all the pulse during the modulation (in a similar manner to~\cite{Fan-reversal,Longhi-reversal}). As a simple estimate, this requires $D \ge N W$ where $W \approx v_g T_p/0.7$ is the FWHM or equivalently,
$$
T_p \le 0.7 D/N v_g,
$$
with typically, $N \ge 3/0.7 \sim 5$. Thus, for example, a few cm-long RM can contain and reverse pulses not longer than a few tens of picoseconds.

The thicknesses of the layers determine the crossing frequency, hence, the carrier frequency, according to Eq.~(\ref{eq:lambda_cr}). Note, however, that the parameter $s$ in Eq.~(\ref{eq:qws_cond}) allows to scale the thickness of one of the layers.

\subsection{Materials} \label{subsec:materials}
\subsubsection{Modulation technique} \label{subsubsec:modulation}
We distinguish between two classes of modulations. First, picoseconds and longer modulation times can be done externally, e.g., by using the electro-optical (Pockels) effect~\cite{Boyd-book} or using carrier injection/depletion in semiconductor materials, see e.g.,~\cite{Lipson_review}. These techniques yield a modulation which is, to leading order, uniform; both can be realized with essentially linear modulators (see also~\cite{Fan-reversal,Longhi-reversal,Lipson_review}).

Large electro-optical coefficients are found in ferroelectric materials such as $LiNbO_3$
, $BaTiO_3$ 
or $BaStTiO_3$, all which can be modulated on the GHz scale and faster. 
Index modulation via carrier injection/depletion can be obtained in a variety of semiconductors. For example, index modulations of the order of $1\%$ at about $100$GHz were achieved in silicon devices~\cite{Lipson_review,Lipson_PINIP}. Even faster modulations of $1$ps can be achieved in ion-implanted silicon~\cite{ps_Si_modulation}. Comparable modulations may be attained in other semiconductors such as GaAs. 


Shorter modulation times can only be achieved using intense ultrashort pump pulses and rely on a nonlinear mechanism such as Cross-Phase modulation (XPM) or carrier injection/depletion~\cite{Lipson_review,Kuipers_Krauss} through a pump-probe type experiment. In such cases, the modulations may not be uniform or even periodic, unless the geometry of the modulation is designed appropriately, see e.g.,~\cite{Miller_OL}.




We note that in the case of XPM-based modulation, the RM structure need not only to have a periodically varying refractive index, but also a periodically varying cubic nonlinear (Kerr) coefficient. A systematic study of pulse propagation in structures with a periodically varying (positive) Kerr coefficient was conducted by Sivan {\em et al.}~\cite{Sivan_PD_06,Sivan_PRL_06}. This work was elaborated and extended to systems in which the magnitude and sign of the Kerr coefficient vary periodically, as well as to structures in which both refractive index and Kerr coefficient vary periodically. For a  recent review of light propagation in such structures, see review by Malomed {\em et al.}~\cite{Boris_review_periodic_kerr}.

\subsubsection{Refractive index}\label{subsubsec:index}
As shown in the formula for the coupling coefficient~(\ref{eq:m_od}), high reversal efficiency can be achieved if (a) there is a high contrast between the refractive indices of both layers and (b) the modulation is as different as possible from such that maintains the QWS condition. Both conditions ensure that a gap as wide as possible is opened by the modulation. In particular, this implies that it is far better to modulate a single layer (e.g., by choosing one of the layers to be air), rather than two layers by the same amount. In this case, the coupling coefficient is given by
\begin{equation}\label{eq:mbf_1layer_nm20}
|m_{od}| = \frac{|n_1 - n_2|}{2 n_j (n_1 + n_2)},
\end{equation}
where $j$ is the index of the modulated layer. In the limit of a high contrast, $m_{od} \to 1/2 n_j$ so that it is clearly advantageous to modulate the {\em low} index layer. If, on the other hand, one modulates the {\em high} index material, the refractive index of the high-index layer required for optimal reversal is $n_{1,opt} = n_2(1 + \sqrt{2})$; still, since in this case the dependence of the reversal coefficient on $n_1$ is rather weak, there is only a small reduction in efficiency for indices slightly different from $n_{1,opt}$.

These configurations can be realized with material combinations such as silicon or $LiNb0_3$ and air, both giving rise to $m_{bf} \sim 0.08$. An alternative is to use a low-index ferroelectric with a high-index material (e.g., a semi-conductor). Then, if only the ferroelectric is modulated, the coupling coefficient can have comparable magnitude. Another possibility is to use high refractive-index materials, either for constructing mid-IR~\cite{MTM_high_index} and THz~\cite{THz_high_index} RMs. The associated efficiencies may be slightly higher. 

Finally, an even higher reversal efficiency can be attained if the refractive index of one layer is increased while the refractive index of the other is decreased. This is possible, for example, with XPM-based modulation in structures in which the Kerr coefficient has different sign in each layer (see~\cite{Boris_review_periodic_kerr} and Section~\ref{subsubsec:modulation}).

\subsection{Material dispersion}\label{subsec:dispersion}
In the derivation of the envelope equations we have ignored the effects of dispersion. These are unavoidable, especially since our design prefers high index materials and long samples. Dispersion affects several aspects of the RM. First, since now the refractive index depends on the wavelength, it is clear that the QWS condition~(\ref{eq:qws_cond}) cannot be satisfied for all wavelengths. However, it can be easily shown using the analytical expressions for the band structure that in order to maintain a zero-gap, it is sufficient for the QWS condition~(\ref{eq:qws_cond}) to be satisfied exactly at the crossing wavelength~(\ref{eq:lambda_cr}).

Second, dispersion will affect the dynamics of the pulses inside the RM. Self-consistent inclusion of dispersion in Maxwell equations requires to compute the dynamics of the Polarization so that the refractive index arises naturally from the constitutive relations~\cite{Taflove-book}. However, for the purpose of the derivation of envelope equations, it is sufficient to introduce the dispersion into the band structure calculation (see~\cite{Yeh-book} or more specifically, Appendix~\ref{app:FB_modes}). In this case, dispersion will cause a deviation from the strictly linear dispersion relations and will be manifested via the addition of high-order time-derivative terms (GVD and higher order dispersion) in each envelope equation. These will lead to broadening and even distortion (in the presence of significant higher-order dispersion) of the pulses, possibly to a different extent on either band. However, since the structural dispersion of the PhC, dominates over material dispersion (which is a weak effect in dielectrics), dispersion is not expected to be significant, especially due to the short lengths of the proposed RMs.

\subsection{Losses}
So far, we have also ignored the possibility of losses in the RM (i.e., absorbance or scattering). As a rough estimate, if one of the layers has non-zero absorbance, then, it is sufficient for the absorption coefficient of the lossy layer, $\alpha = 2 k_0 n_j''$, to satisfy $\alpha \frac{d_1}{d} D \ll 1$. For example, typical losses in silicon nanostructures (e.g., channel or ridge waveguides where losses originate mainly from scattering from the roughness of the interfaces) vary between $0.05 - 1/cm$~\cite{Lipson_review}. For a silicon-air QWS ($n_{Si} \sim 3.5$, so that $d \sim (n_1 + n_2) d_1 / n_2 \sim 4.5 d_1$), this corresponds to the condition $\alpha \ll \frac{d}{d_1 D} \sim \frac{4.5}{D}$. Thus, such losses can be sufficiently low even for a few cm-long RM. If the material of choice has even higher losses, one can choose a different value of $s$ in order to reduce the thickness of the absorbing layer. Thus, overall, losses are not expected to cause any significant reduction in performance of our RM.

\section{Numerics}\label{sec:numerics}
In order to validate the analysis, we have performed extensive numerical simulations. First, the wave equation~(\ref{eq:waeq}) was solved with standard absorbing boundary conditions at both ends of the domain, namely,
\begin{eqnarray} \nonumber
E_x(x = 0,t) &=& \frac{n(x = 0,t)}{c} \left(E_t - 2E_{inc,t}\right), \nonumber \\
E_x(x = D,t) &=& - \frac{n(x = D,t)}{c} E_t. \nonumber
\end{eqnarray}
Second, the envelope equations~(\ref{eq:f_uniform})-(\ref{eq:b_uniform}) were solved, again, with standard absorbing boundary conditions for both envelopes with only the forward envelope initially excited, namely,
\begin{eqnarray}
f_x(x = 0,t) &=& \frac{n(x = 0,t)}{c} \left(f_t - 2f_{inc,t}\right), \nonumber \\
f_x(x = D,t) &=& - \frac{n(x = D,t)}{c} f_t, \nonumber \\
b_x(x = 0,t) &=& \frac{n(x = 0,t)}{c} b_t, \nonumber \\
b_x(x = D,t) &=& - \frac{n(x = D,t)}{c} b_t. \nonumber
\end{eqnarray}

Although the numerical solution of these equations does not require to resolve the $T$- (and $\lambda$-) scale oscillations, it still poses a non-trivial challenge. First, note that since the modulation is uniform in space, then, unlike the case of light propagation through nonlinear Kerr media~\cite{shallow_grating_review,deep_gratings_PRE_96} where the modulation is caused by the moving pulse itself, Eqs.~(\ref{eq:f_uniform})-(\ref{eq:b_uniform}) cannot be reduced to a set of two coupled {\em ordinary} differential equations. For this reason, simple numerical schemes such as the one described in~\cite{CMT_counter_prop_numerics} are not applicable to the current problem and one must employ partial differential equations techniques. Second, the absence of diffusion from the transport equations~(\ref{eq:f_uniform})-(\ref{eq:b_uniform}) makes any discretization scheme unstable. In order to retain numerical stability, one can add small artificial diffusion, or alternatively, use the Galerkin-Ritz technique which is relatively non-diffusive.

\begin{figure}[htbp]
\centering{\includegraphics[scale=0.95]{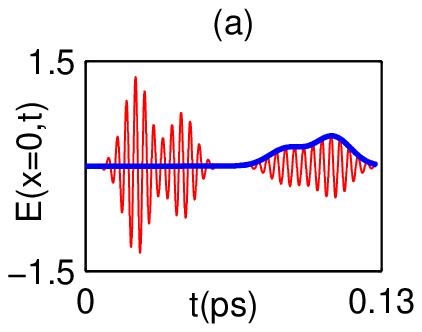} 
\includegraphics[scale=0.95]{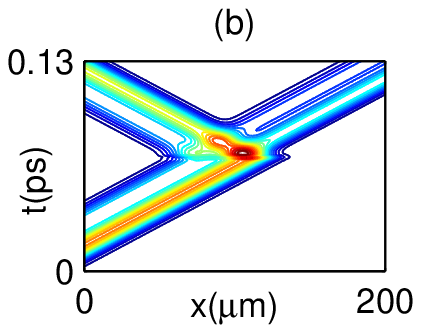}} 
\caption[]{(Color online) (a) Amplitude of an asymmetric pulse (Eq.~(\ref{eq:waeq}); red line) and the associated backward wave envelope (Eq.~(\ref{eq:b_uniform}); blue line) at the input side of the QWS as a function of time. (b) A spatio-temporal contour map of forward and backward envelopes in (a). } 	\label{fig:rev_illustration}
\end{figure}

\begin{figure}[htbp]
\centering{\includegraphics[scale=1.00]{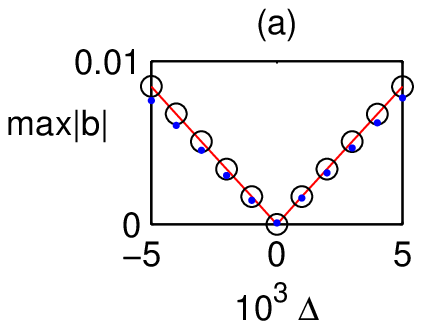} \includegraphics[scale=1.00]{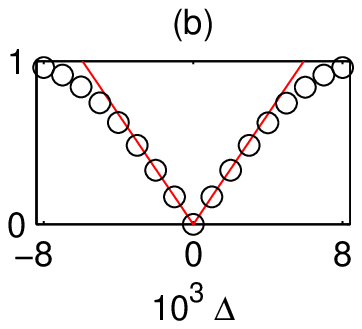}} 
\caption[]{(Color online) (a) Reversal of a $T_p = 100$fs, unit-amplitude Gaussian input pulse~(\ref{eq:f_inc_gaussian}) under a Gaussian modulation~(\ref{eq:gaussian_mod}) as a function of the index change $\Delta$ for $n_1 = 3.45$, $n_2 = 1$, $\lambda_c = 1550$nm, and $T_{mod} = 10$fs. Shown are numerical solutions of the wave equation~(Eq.~(\ref{eq:waeq}); blue dots) vs. the solution of the envelope equations~(Eq.~(\ref{eq:b_uniform}); black circles) and the analytical solution~(Eq.~(\ref{eq:abs_b_weak_small_m0}); red solid line). (b) Same as (a) for $T_{mod} = 1$ps and $T_p = 10$ps. } 	\label{fig:rev_efficiency}
\end{figure}

In Fig.~\ref{fig:rev_illustration}(a), we plot the pulse profile at the input side of the PhC as a function of time. The reversed pulse has a somewhat lower-amplitude, but the leading and trailing edges have clearly exchanged roles, so that wave-front has been accurately reversed. We also show that the solutions of the wave equation~(\ref{eq:waeq}) and the envelope equations~(\ref{eq:f_uniform})-(\ref{eq:b_uniform}) are in excellent agreement. Fig.~\ref{fig:rev_illustration}(b) shows a spatio-temporal contour map of the pulse propagation. It shows that the modulation, occurring once the pulse is in the middle of the RM, causes a fast and complicated dynamics after which the pulse splits into a reversed and a (somewhat delayed) forward component. Note that the modulation causes a delay of the forward propagating pulse (occurring even for very small modulations). This effect is irrelevant to the performance of the RM, and is not captured by the analysis performed in this article.

A comparison of the reversal efficiencies as a function of the modulation strength, which also includes the analytical approximation~(\ref{eq:abs_b_weak_small_m0}), is shown in Fig.~\ref{fig:rev_efficiency}. We employ realistic parameters corresponding to a Silicon-air QWS with $\lambda_c = 1550$nm; such a PhC can be fabricated on-chip with current technology~\cite{Lipson_review}. First, we perform simulations of $\sim20$-cycle pulses. 
Fig.~\ref{fig:rev_efficiency}(a) verifies that all three solutions are in good agreement, thus, validating the analysis. The reversal efficiencies in this case are rather low, however, they are comparable to efficiencies of nonlinear conversion processes of such short pulses. A possible solution is simply to amplify the signal after the reversal process.

In Fig.~\ref{fig:rev_efficiency}(b) we show simulations for longer pulses for which memory and running time required for the solution of Eq.~(\ref{eq:waeq}) are beyond standard available computing resources. Thus, we only show the solutions of the envelope equations~(\ref{eq:f_uniform})-(\ref{eq:b_uniform}) vs. the analytical solution. As in Fig.~\ref{fig:rev_efficiency}(a), there is very good agreement between the numerical and analytical solutions up to high efficiencies ($\sim 50\%$). At even higher efficiencies, the analytical approximation overestimates the reversal efficiency as given by the solution of the envelope equations. This is because at such high efficiencies, the forward wave amplitude is significantly decreased, so that the forward-to-backward wave coupling decreases as well. Nevertheless, the solutions of the envelope equations show that a $100\%$ reversal efficiency can be achieved in this configuration with index changes only slightly higher than those predicted analytically. Thus, our RM has comparable performance to the previously suggested time-reversal schemes in index-modulated coupled-resonator arrays of optical waveguides~\cite{Fan-reversal,Longhi-reversal}. However, in contrast to these schemes, the QWS admits pulses of very broad spectrum. In addition, our system does not suffer from a deterioration of performance due to non-adiabatic modulations.




\section{Acknowledgements}
Y.S. would like to thank A.I. Fernand\'ez-Domingu\'ez and A. Aubry for innumerable valuable discussions. The research of Y.S. was supported by the Royal Society Newton Fellowship. J.B.P was supported by the EU project PHOME (Contract No. 213390).

\appendix
\section{Floquet-Bloch modes near the crossing point}\label{app:FB_modes}
The Floquet-Bloch modes are solutions of the Helmholtz equation, obtained by Fourier transforming the wave equation~(\ref{eq:waeq}) with $M_0 = 0$. In a single unit cell, they are given by 
\begin{eqnarray}\label{eq:FB_modes}
\phi(x) &=& \frac{1}{N_\phi} e^{i \kappa x} u(x), \nonumber \\ 
u &=& e^{- i \kappa x} \left\{
\begin{array}{ll}
a_0 e^{i k_1 x} + b_0 e^{- i k_1 x}, & 0 < x < d_1, \\
c_0 e^{i k_2 (x - d_1)} + d_0 e^{- i k_2 (x - d_1)}, & d_1 < x < d,
\end{array} \nonumber 
\right.
\end{eqnarray}
where $k_j = \omega n_j(\omega) / c$, $N_\phi$ is a normalization constant chosen such that
\begin{equation}\label{eq:orthogonality_phi}
\int_0^d \phi_a^*(x) n^2(x) \phi_b(x) dx = \delta_{ab},
\end{equation}
and the FB wavenumber is given by
\begin{eqnarray}\label{eq:disp_relations}
\kappa(\omega) = \frac{1}{d} acos\left(Re\left[A\right]\right),
\end{eqnarray}
with $A$ being the first element of the transfer matrix of the layered system, see~\cite{Yeh-book} and Eq.~(\ref{eq:b0}) below. Note that for simplicity of notation, we omit the $\kappa$ subscript from $u$, $\phi$ and $\Psi$.

For normal incidence, 
the coefficients for the first layer are given by~\cite[Eq.~(6.2-7)]{Yeh-book}
\begin{subequations}
  \label{eq:a0b0}
\begin{eqnarray}
a_0 &\equiv& B = - e^{i k_1 d_1} \frac{i}{2}\frac{n_2^2 - n_1^2}{n_1 n_2} \sin(k_2 d_2), \label{eq:a0} \\
b_0 &\equiv& e^{- i \kappa d} - A, \\ 
A &=& e^{- i k_1 d_1}\left[\cos(k_2 d_2) - \frac{i}{2}\frac{n_2^2 + n_1^2}{n_1 n_2} \sin(k_2 d_2)\right]. \label{eq:b0}
\end{eqnarray}
\end{subequations}
The coefficients for the second layer are given by~\cite[Eq.~(6.1-11)]{Yeh-book}
\begin{subequations}
  \label{eq:c0d0}
\begin{eqnarray}
c_0 &=& \frac{1}{2} \left[e^{- i k_1 d_1} \left(1 + \frac{n_1}{n_2}\right) a_0 + e^{i k_1 d_1} \left(1 - \frac{n_1}{n_2}\right) b_0\right], \nonumber \\ \\
d_0 &=& \frac{1}{2} \left[e^{- i k_1 d_1} \left(1 - \frac{n_1}{n_2}\right) a_0 + e^{i k_1 d_1} \left(1 + \frac{n_1}{n_2}\right) b_0\right]. \nonumber \\
\end{eqnarray}
\end{subequations}

By Eq.~(\ref{eq:lambda_cr}), it follows that
$k_1 d_1 = \pi - \frac{n_1 d_1}{c} \delta\omega = k_2 d_2 /s = \pi - \frac{n_2 d_2}{s c} \delta\omega$, where $\delta\omega = \delta\omega_{f/b}$ (see Eq.~(\ref{eq:d_omega})). 
Accordingly, to $O(\delta\omega^2)$ accuracy, {\em for any value of $s$} we get
\begin{eqnarray}
A &\cong& 1 + i \frac{n_1 d_1}{c} \delta\omega \frac{(n_2 + n_1)^2}{2 n_1 n_2} - \left(\frac{n_1 d_1}{c} \delta\omega\right)^2 \frac{(n_2 + n_1)^2}{2n_1 n_2}, \nonumber
\end{eqnarray}
so that by Eq.~(\ref{eq:disp_relations}),
\begin{eqnarray}
\cos\left(\kappa(\omega) d\right) \cong 1 - \left(\frac{n_1 d_1}{c} \delta\omega\right)^2 \frac{(n_2 + n_1)^2}{2n_1 n_2}. \nonumber
\end{eqnarray}
Since we are interested in a transition between modes at a given (say positive) $\kappa$, it follows that
\begin{eqnarray}\label{eq:kappa}
\kappa(\omega) \cong \frac{\sqrt{|n_1 n_2|}}{c} |\delta\omega| > 0,
\end{eqnarray}
where the absolute value over the frequency ensures that the modes under consideration have the same Bloch momentum, as required for a vertical transition, and the absolute value over the indices ensures that the group-velocity is real even for negative index materials. For simplicity, in what follows we drop the absolute values.

Using all the above, it can also be shown that, to $O(\delta\omega^2)$ accuracy, the coefficients are given approximately by
\begin{eqnarray}\label{eq:ab_FB} 
a_0^{(f,b)} &\cong& i \gamma^{(f,b)} (n_2 - n_1), \nonumber \\
b_0^{(f,b)} &\cong& - i \gamma^{(f,b)} \left(\sqrt{|n_2|} \mp \sqrt{|n_1|}\right)^2,
\end{eqnarray}
where $\gamma^{(f,b)} \equiv \frac{d}{2 c} \delta\omega^{(f,b)}$ is a real, dimensionless, arbitrarily small parameter that represents the detuning from the crossing point.

By Eqs.~(\ref{eq:c0d0}), the coefficients for the second layer are given by
\begin{eqnarray} \label{eq:cd_FB}
c_0^{(f,b)} &=& \mp \sqrt{\frac{n_1}{n_2}} a_0^{(f,b)}, \quad d_0^{(f,b)} = \pm \sqrt{\frac{n_1}{n_2}} b_0^{(f,b)}.
\end{eqnarray}
These relations show that $A \to 1^-$ at the crossing point, in which the mode is identically zero. Thus, the gap indeed consists only of the crossing point.

It follows that
\begin{eqnarray}\label{eq:phi_sim_int1}
N_{f/b}^2 \int_0^{d_1} |\phi_{f/b}|^2 \cong 2 d_1 \gamma^2 (n_2 + n_1)\left(\sqrt{n_2} \mp \sqrt{n_1}\right)^2, \nonumber
\end{eqnarray}
and similarly,
\begin{eqnarray}\label{eq:phi_sim_int2}
N_{f/b}^2 \int_{d_1}^d |\phi_{f/b}|^2 &\cong& 2 d_2 \gamma^2 \frac{n_1}{n_2} (n_2 + n_1) \left(\sqrt{n_2} \mp \sqrt{n_1}\right)^2. \nonumber
\end{eqnarray}
Hence, the normalization constants are
\begin{equation}\label{eq:Nfb}
N_{f/b} \equiv \int_0^d n^2_{PhC}(x) |\phi_{f/b}|^2 = 2 n_1 \gamma \sqrt{d_1 (n_2 + n_1)}\left(\sqrt{n_2} \mp \sqrt{n_1}\right),
\end{equation}
and the mode~(\ref{eq:FB_modes}) in the first layer is given by 
\begin{eqnarray}\label{eq:FB_modes_final}
\phi_{f/b}(0 < x < d_1) 
\sim \left[\frac{n_2 - n_1}{\sqrt{n_2} \mp \sqrt{n_1}} e^{i k_1 x} - \left(\sqrt{n_2} \mp \sqrt{n_1}\right) e^{- i k_1 x} \right], \nonumber
\end{eqnarray}
with similar expressions for the second layer. Note that since for dielectric materials, $n_2 + n_1 > \sqrt{n_2} + \sqrt{n_1} \gg n_2 - n_1 > \sqrt{n_2} - \sqrt{n_1}$, the forward (backward) plane wave component is dominant for the forward wave envelope $f$ (backward wave envelope $b$), respectively. Note that the size of the coefficient is independent of the frequency, at least for non-dispersive materials, so that the dominance of the forward/backward component is abruptly switched across the zero gap.

The elements of the uni-directional FB modes~(\ref{eq:leading_order}) are related to the (standard) FB modes~(\ref{eq:FB_modes}) through~\cite{deep_gratings_PRE_96}
\begin{equation}\label{eq:psi_pm}
\psi^\pm(x_0) = \frac{1}{2}\left(\sqrt{n_{PhC}(x_0)}\phi(x_0) \mp \frac{i c}{\omega \sqrt{n_{PhC}(x_0)}} \frac{\partial \phi}{\partial x_0}\right).
\end{equation}
It is straight-forward to verify that $\psi^\pm$ satisfy a normalization relation similar to~(\ref{eq:orthogonality_phi}).

\section{Calculation of coefficients in envelope equations}\label{app:CMT_coeffs}
In this section we compute the coefficients of the envelope equations~(\ref{eq:v_ab})-(\ref{eq:m1_ab}) using the approximate analytical expressions obtained for the FB modes in Appendix~\ref{app:FB_modes}. In particular, below we rely on Eqs.~(\ref{eq:FB_modes}), (\ref{eq:ab_FB})-(\ref{eq:cd_FB}) and (\ref{eq:Nfb})-(\ref{eq:psi_pm}). All the analytical expressions derived below were found to be in perfect agreement with numerical evaluation of the definitions~(\ref{eq:v_ab})-(\ref{eq:m1_ab}).

\subsection{$v$}\label{app:v}
By Eq.~(\ref{eq:v_ab}) and Eq.~(\ref{eq:psi_pm}), it follows that
\begin{eqnarray}
v_{ab} = - \frac{c^2}{2} Im \int_0^d  \left[\frac{1}{\omega_a} \phi_b \frac{\partial \phi_a^*}{\partial x_0} - \frac{1}{\omega_b} \phi_a^* \frac{\partial \phi_b}{\partial x_0}\right] dx_0. \nonumber
\end{eqnarray}
The diagonal terms reduce to
\begin{eqnarray}
N_a^2 v_{aa} \cong n_1 d c \left(|a_0^{(a)}|^2 - |b^{(a)}_0|^2\right). \nonumber
\end{eqnarray}
Using Eq.~(\ref{eq:ab_FB}) gives
\begin{eqnarray}
N_a^2 v_{aa} \cong \pm 4 n_1 d c \gamma^2 \sqrt{n_2 n_1} \left(\sqrt{n_2} \mp \sqrt{n_1}\right)^2, \nonumber
\end{eqnarray}
from which it follows that
\begin{eqnarray}
v_{aa} \cong \pm \frac{c}{\sqrt{n_1 n_2}}.
\end{eqnarray}
Thus, by Eq.~(\ref{eq:kappa}), we obtain that $v_{ff} = - v_{bb} = v_g$. 

Similarly, one can show that the off-diagonal terms vanish, $v_{ab} = 0$ so that, indeed, there is no reflection-coupling in the system.

\subsection{$m^{(0)}$}\label{app:m0}
By Eq.~(\ref{eq:m0_ab}) and Eq.~(\ref{eq:psi_pm}), it follows that 
\begin{eqnarray}
m^{(0)}_{ab} &=& \int_0^d \underline{\Psi}^\dagger_a(x_0) p \underline{\Psi}_b(x_0) dx_0 = \int_0^d p(x_0) \left[{\psi^+_a}^* \psi^+_b + {\psi^-_a}^* \psi^-_b\right] dx_0 \nonumber \\
&=& \frac{1}{2} \int_0^d p(x_0) n_{PhC}(x_0) \left(\phi_a^* \phi_b + \frac{c^2}{\omega_a \omega_b n_{PhC}^2} \frac{\partial \phi_a^*}{\partial x_0} \frac{\partial \phi_b}{\partial x_0}\right) dx_0. \nonumber
\end{eqnarray}
Then, it can be shown that 
\begin{widetext}\begin{eqnarray}\label{eq:m_ab_temp}
m^{(0)}_{ab} &=& \int_0^d p(x_0) n_{PhC}(x_0) \phi_a^* \phi_b dx_0 = \frac{\left({a_0^{(a)}}^* a_0^{(b)} + {b_0^{(a)}}^* b_0^{(b)}\right) n_1 \bar{n}_{m1} d_1 + \left({c_0^{(a)}}^* c_0^{(b)} + {d_0^{(a)}}^* d_0^{(b)}\right) n_2 \bar{n}_{m2} d_2}{N_a N_b} \\
&+& \frac{n_1}{N_a N_b} \int_0^{d_1} p(x_0) \left({a_0^{(a)}}^* b_0^{(b)} e^{- 2 i k_1 x_0} + {b_0^{(a)}}^* a_0^{(b)} e^{2 i k_1 x_0}\right) dx_0 + \frac{n_2}{N_a N_b} \int_{d_1}^d p(x_0) \left({c_0^{(a)}}^* d_0^{(b)} e^{- 2 i k_1 x_0} + {d_0^{(a)}}^* c_0^{(b)} e^{2 i k_1 x_0}\right) dx_0. \nonumber
\end{eqnarray}
In Eq.~(\ref{eq:m_ab_temp}), the first term gives the contribution of the zero Fourier component of the modulation and the last two terms give the contribution of the $2k_0 n_j$ Fourier component of the modulation.

Let us now separate into the diagonal and off-diagonal cases. In the former case, it can be shown that
\begin{eqnarray}\label{eq:m0_aa_final}
m^{(0)}_{aa} &=& \frac{\bar{n}_{m1}}{2 n_1} + \frac{\bar{n}_{m2}}{2 n_2} - \frac{n_2 - n_1}{2 n_1 d_1 (n_2 + n_1)} \left(\int_0^{d_1} p(x_0) \cos\left(2 k_1 x_0\right) dx - \int_{d_1}^d p(x_0) \cos\left(2 k_2 x_0\right) dx_0\right). 
\end{eqnarray}
Thus, Eq.~(\ref{eq:m0_aa_final}) shows that both diagonal coefficients are identical and that it is only the $2 k_0 n_j$ Fourier components of $p(x_0)$ which contribute to the reversal.

For a piecewise-uniform modulation, the latter contribution vanishes so that
\begin{eqnarray}
m^{(0)}_{ff} = m^{(0)}_{bb} = \frac{n_{m1}}{2 n_1} + \frac{n_{m2}}{2 n_2}.
\end{eqnarray}

For off-diagonal coefficients, it can be shown that
\begin{eqnarray}\label{eq:?}
m^{(0)}_{ab} &\cong& \frac{n_2 - n_1}{n_2 + n_1} \frac{\bar{n}_{m1} d_1 - \bar{n}_{m2} d_2}{2 n_1 d_1} - \frac{\int_0^d p(x_0) \left[(n_2 \mp n_1)^2_b e^{- 2 i k_0 n_{PhC}(x_0) x_0} + (n_2 \mp n_1)^2_a e^{2 i k_0 n_{PhC}(x_0) x_0}\right] dx_0}{{{4 n_1 d_1 (n_2 + n_1)}}}. \nonumber \\
\end{eqnarray}
For a piecewise-uniform modulation, the contribution of the integrals vanishes so that
\begin{eqnarray}
m^{(0)}_{fb} &=& \frac{n_2 - n_1}{2(n_2 + n_1)} \left(\frac{n_{m1}}{n_1} - \frac{n_{m2}}{n_2}\right).
\end{eqnarray}

\subsection{$m^{(1)}$}\label{app:m1}
By Eq.~(\ref{eq:m1_ab}),
\begin{eqnarray}
m^{(1)}_{ab} &=& \int_0^d \underline{\Psi}^\dagger_a(x_0) p \underline{\underline{C}}^T \underline{\Psi}_b(x_0) dx_0 = m^{(0)}_{ab} + \frac{1}{2}\int_0^d p(x_0) \left[{\psi^+_a}^* \psi^-_b + {\psi^-_a}^* \psi^+_b\right] dx_0. \nonumber 
\end{eqnarray}
This can be shown to equal 
\begin{eqnarray}
m^{(1)}_{ab} &=& m^{(0)}_{ab} + \frac{1}{2} \int_0^d p(x_0) n(x_0) \phi_a^* \phi_b dx_0 - \frac{c^2}{2 \omega_a \omega_b} \int_0^d \frac{p(x_0)}{n_{PhC}(x_0)} \frac{\partial \phi_a^*}{\partial x_0} \frac{\partial \phi_b}{\partial x_0} dx_0. \nonumber
\end{eqnarray}
Then, using Eqs.~(\ref{eq:ab_FB})-(\ref{eq:cd_FB}), it can be shown that
\begin{eqnarray}
m^{(1)}_{ab} &=& m^{(0)}_{ab} + \frac{n_1}{N_a N_b} \int_0^{d_1} p(x_0) \left({a_0^{(a)}}^* b_0^{(b)} e^{- 2 i k_1 x_0} + {b_0^{(a)}}^* a_0^{(b)} e^{2 i k_1 x_0}\right) dx_0 \nonumber \\
&+& \frac{n_2}{N_a N_b} \int_{d_1}^d p(x_0) \left({c_0^{(a)}}^* d_0^{(b)} e^{- 2 i k_1 x_0} + {d_0^{(a)}}^* c_0^{(b)} e^{2 i k_1 x_0}\right) dx_0.
\end{eqnarray}
It follows that the diagonal terms are identical and that for a piecewise-uniform modulation, $m^{(1)}_{ab} \equiv m^{(0)}_{ab}$, i.e., the off-diagonal elements in $C_T$ do not contribute to the dynamics. 
\end{widetext}


\end{document}